\newcommand\norm[1]{\lVert#1\rVert}
\documentclass{article}
\usepackage{spconf,amsmath,amsfonts,graphicx}
\usepackage{subcaption}
\PassOptionsToPackage{hyphens}{url}\usepackage{hyperref}
\usepackage{multirow}
\usepackage{booktabs}

\usepackage[absolute,showboxes]{textpos}

\TPMargin{5pt}

\newcommand{\copyrightstatement}{
    \begin{textblock}{0.84}(0.08,0.93)    % tweak here: {box width}(leftposition, rightposition)
         \noindent
         \footnotesize
         \copyright  2024 IEEE. Personal use of this material is permitted. Permission from IEEE must be obtained for all other uses, in any current or future media, including reprinting/republishing this material for advertising or promotional purposes, creating new collective works, for resale or redistribution to servers or lists, or reuse of any copyrighted component of this work in other works.
    \end{textblock}
}

\title{Microphone Conversion: Mitigating Device Variability in Sound Event Classification}
\name{$^{\star}$Myeonghoon Ryu$^{1, 2}$\thanks{$^{\star}$These authors contributed equally to this work.} \qquad $^{\star}$Hongseok Oh$^{1, 3}$ \qquad Suji Lee$^{1}$ \qquad Han Park$^{1}$}
\address{
    $^{1}$ Deeply Inc.\\
    $^{2}$ Seoul National University \\
    $^{3}$ University of California, San Diego
}

\begin{document}

\copyrightstatement
\maketitle
\begin{abstract}
In this study, we introduce a new augmentation technique to enhance the resilience of sound event classification (SEC) systems against device variability through the use of CycleGAN. We also present a unique dataset to evaluate this method. As SEC systems become increasingly common, it is crucial that they work well with audio from diverse recording devices. Our method addresses limited device diversity in training data by enabling unpaired training to transform input spectrograms as if they are recorded on a different device. Our experiments show that our approach outperforms existing methods in generalization by 5.2 - 11.5\% in weighted f1 score. Additionally, it surpasses the current methods in adaptability across diverse recording devices by achieving a 6.5\% - 12.8\% improvement in weighted f1 score.
\end{abstract}

\section{Introduction}
\label{sec:intro}
\begin{figure*}[t]
  \centering
  \includegraphics[width=\linewidth]{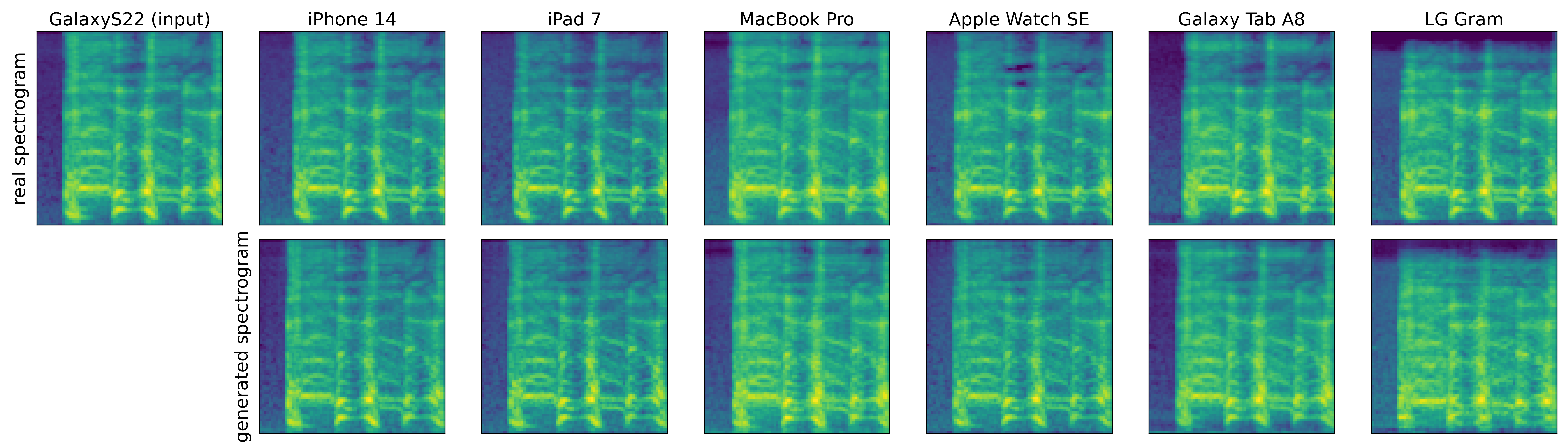}
  \caption{Spectrograms of the real and generated coughing sounds are presented in the top and bottom rows, respectively. The generated ones are produced by corresponding Microphone Conversion networks using a real spectrogram of Galaxy S22.}
  \label{fig:real_gen_spectrogram}
\end{figure*}

Sound event classification (SEC) aims to automatically identify various types of sounds, like speech, music, and environmental sound, using signal processing and machine learning techniques. Despite recent progress that has led  to practical applications, existing models still struggle with distortions caused by different recording devices. These distortions, although often subtle to human ears as shown in Figure \ref{fig:real_gen_spectrogram}, can significantly hamper SEC system performance\cite{Martínmorato2022lowcomplexity}.

Past strategies to tackle this domain shift have largely centered on data augmentation and normalization. Initiatives like the DCASE Challenge\cite{Martínmorato2022lowcomplexity, Turpault2019_DCASE} have tackled this problem but have limitations due to their focus on synthetic evaluation data, reducing the thoroughness of its evaluations.

Previous studies have applied CycleGAN to tasks like speaker verification and emotion recognition\cite{mathur2019mic2mic, kataria2021deep}. However, they depended on datasets lacking detailed device information and did not use optimal recording environments. This oversight can introduce biases and challenges.

To bridge these gaps, we created a dataset featuring sound events recorded in an anechoic chamber using real-world devices, simultaneously. This dataset facilitates a deeper understanding of device-induced performance variations. 

Our key contribution is the introduction of the unique dataset and Microphone Conversion, a plug-and-play augmentation technique for SEC systems. With the CycleGAN\cite{zhu2017unpaired} framework, our method enables more effective deployment on heterogeneous devices without requiring paired or labeled data. This plug-and-play solution generates spectrograms resembling recordings from desired devices. It empowers SEC systems to either generalize across devices or optimize for a specific one, showing enhanced performance compared to existing state-of-the-art methods.

\section{Dataset}
In this section, we outline the process of creating the unique dataset and provide an overview of its contents.

\subsection{Sound Events}
\label{section:sound_event}
The dataset contains 75 types of sound. Of these, 25 were directly produced and recorded, while the remaining 50 are recordings of the playback of anechoic recordings from the RWCP Sound Scene Database (RWCP-SSD)\cite{nakamura2000acoustical}. The former includes sound of things, human sound, and musical instruments. The RWCP-SSD contains non-speech sounds recorded in an anechoic room. These RWCP-SSD samples were played using a KRK VXT6 loudspeaker. We manually annotated the start and end times of each sound event, resulting in a paired dataset intended for fair evaluation.

\subsection{Devices}
\label{section:devices}
We have constructed our dataset using 18 distinct recording devices. Of these, 11 end-user devices come with built-in microphones, chosen specifically to capture distortions caused by varying microphones and digital signal processing units. Additionally, we have included 7 microphones to investigate the distortions due to transducer types and polar patterns.

\subsection{Environment}
All recordings were made in an anechoic chamber meeting ISO 3745 standards\footnote{https://www.iso.org/standard/45362.html}. Sound sources were located on a side of the chamber, while the recording devices were set 1.5 meters away, with their microphones aimed at the sources. 

\section{Our Method}
The objective of Microphone Conversion is to train a mapping function  $F$ that transforms source device spectrograms $X_A$ into those of a target device. In essence, the generator produces output spectrograms  $F(X_A)$ that are statistically similar to the data distribution of the target device B.

\subsection{CycleGAN}
Our approach employs CycleGAN, an unsupervised framework for image-to-image translation that can learn mappings between two domains with unpaired data. Built on the Generative Adversarial Networks (GANs), CycleGAN involves two generators, $F$ and $G$, and two discriminators, $D_A$ and $D_B$. 

\begin{align}
    \label{eq:gan_loss}
    \mathcal{L}_{adv}(F, D_B, X_A, X_B) \nonumber
    &= \mathbb{E}_{X_A{\sim}p(x_a)}[{\log}(1-D_B(F(X_A)))] \nonumber \\ 
    &+ \mathbb{E}_{X_B{\sim}p(x_b)}[{\log}D_B(X_B)]
\end{align}
\begin{align}
    \label{eq:cycle_loss}
    \mathcal{L}_{cycle}(F, G, X_A, &X_B) \nonumber
    = \mathbb{E}_{X_A{\sim}p(x_a)}[\norm{G(F(X_A)) - X_A}_{1}] \nonumber \\ 
    &+ \mathbb{E}_{X_B{\sim}p(x_b)}[\norm{F(G(X_B)) - X_B}_{1}]
\end{align}

Each generator performs bijective mapping between domain A and B, serving as inverse functions of one another. The discriminators aim to differentiate real images from converted counterparts. The adversarial loss (Eq. \ref{eq:gan_loss}) encourages the generators to produce outputs indistinguishable from the target domain, while the cycle-consistency loss (Eq. \ref{eq:cycle_loss}) ensures that the generated images, when reverse-mapped, closely resemble their originals. This dual-loss setup allows CycleGAN to preserve the content while translating the style. 

\begin{figure}[t]
    \centering 
    \includegraphics[width=\linewidth]{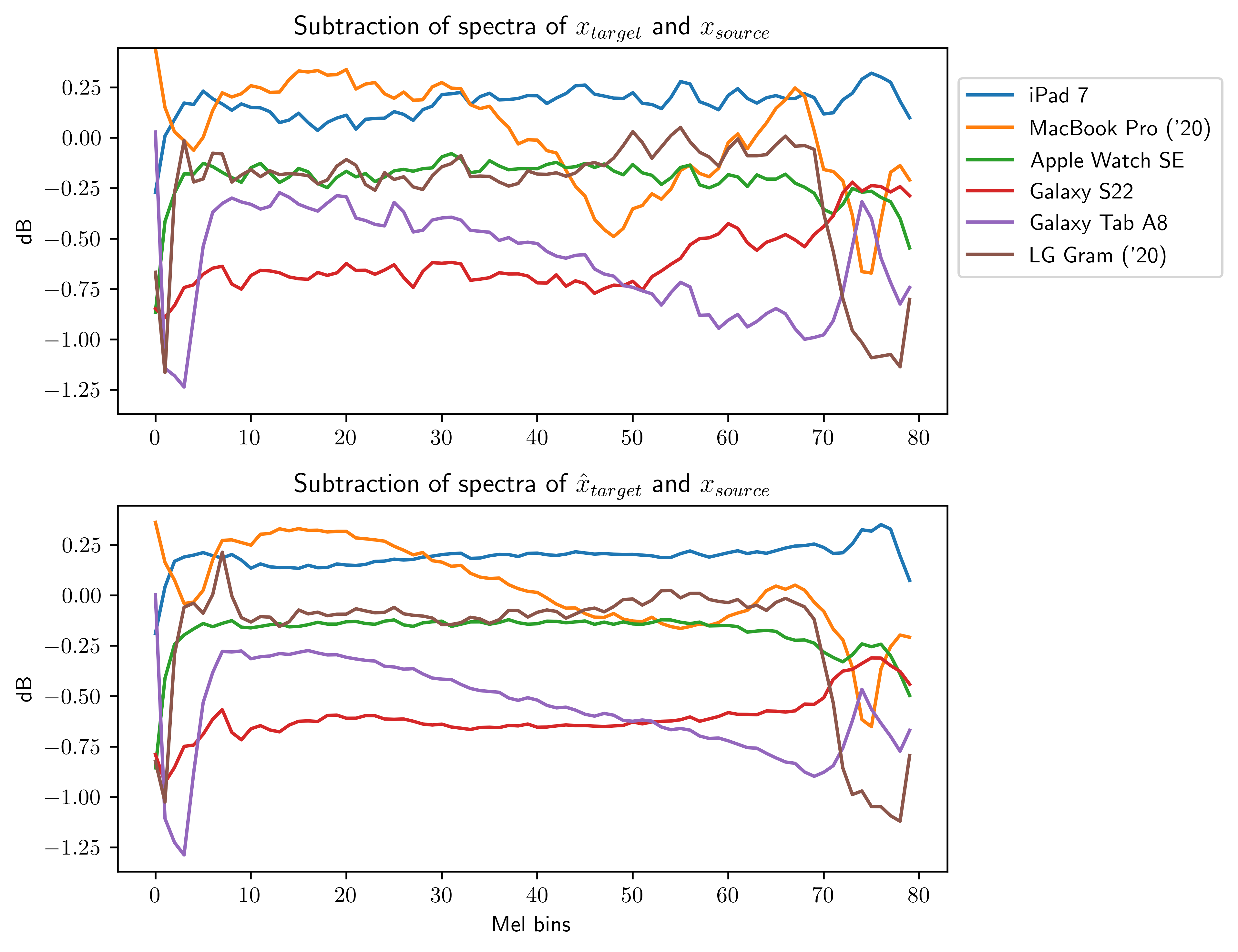}
    \caption{A difference spectra between a real spectrum (${x}_{source}$, iPhone 14) and a real/generated spectrum (${x}_{target}$, $\hat{x}_{target}$, Other devices). Spectra are calculated using Welch's method.}
    \label{fig:impulse_response}
\end{figure}

\subsection{Development Data}
\label{section:dataset}
To build a development set for our experiment, we select data from seven end-user devices from \emph{$data_{full}$}. We create 315,966 audio segments using a 930-ms window and a 50\% overlap. They are filtered based on a 10\% threshold for sparse classes like `coughing' and `clap,' and a 50\% threshold for the others, resulting in a 246,544-segment development set.

The entire development set is then divided into three subsets: \emph{$data_{train,mc}$}, \emph{$data_{train,sec}$}, and \emph{$data_{val}$} are 45\%, 45\%, 10\%, respectively. We use stratification to balance sound events across subsets and to guarantee that each audio segment has counterparts from different devices in each subset. 

\begin{figure}[t]

\begin{minipage}[b]{.48\linewidth}
  \centering
  \centerline{\includegraphics[width=\linewidth]{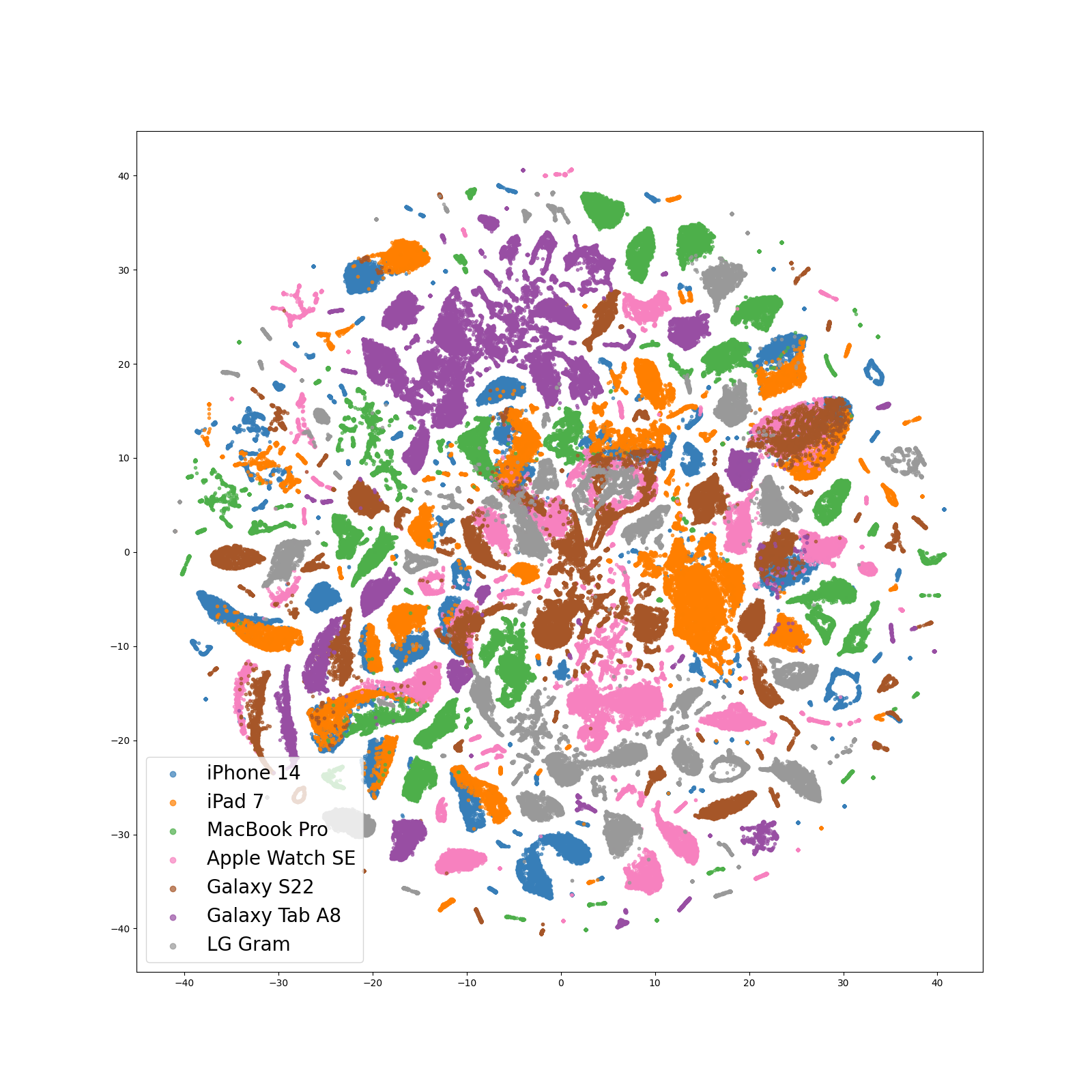}}
  \subcaption{All sound}
  \label{fig:tsne(a)}
\end{minipage}
\hfill
\begin{minipage}[b]{0.48\linewidth}
  \centering
  \centerline{\includegraphics[width=\linewidth]{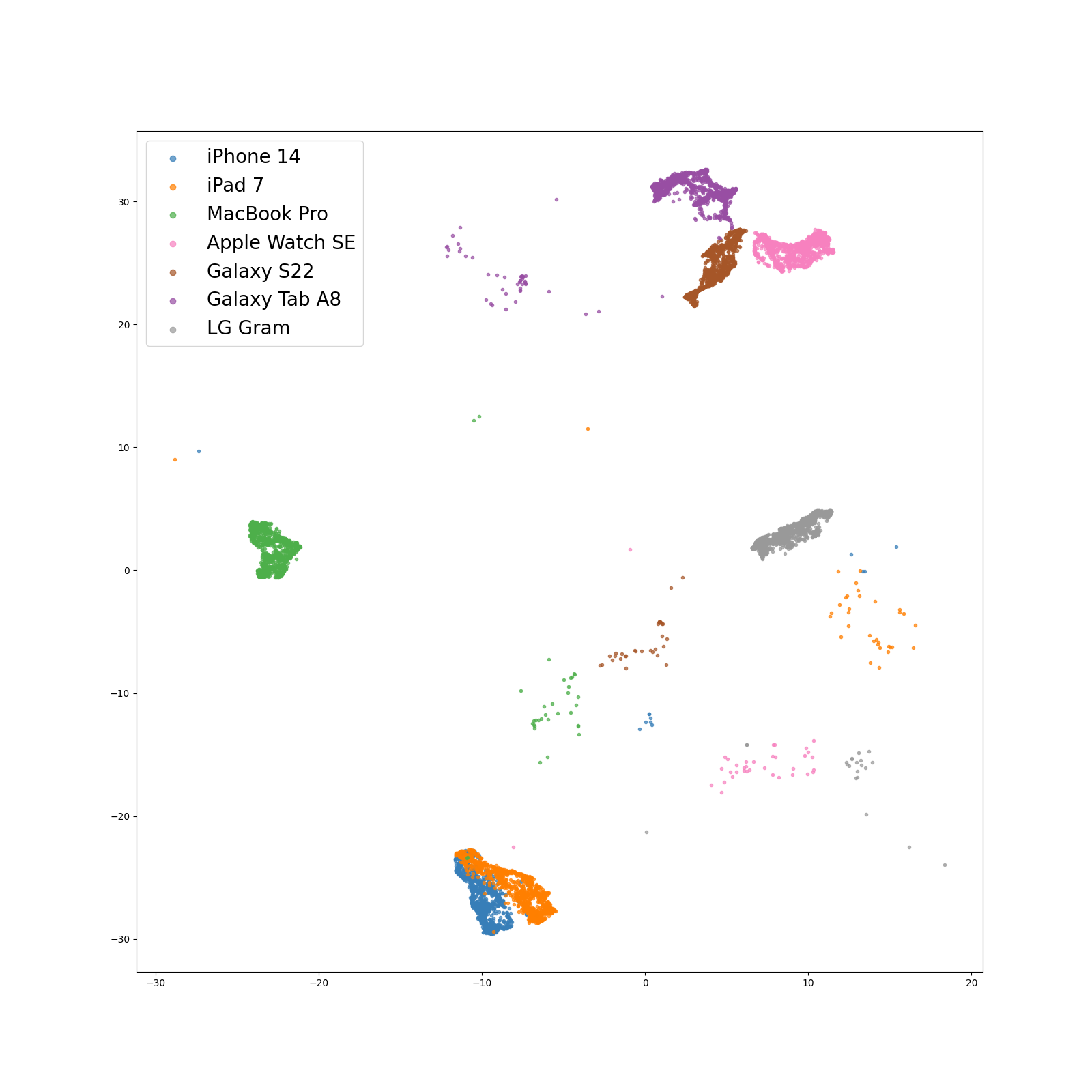}}
  \subcaption{Whistle}
  \label{fig:tsne(b)}
\end{minipage}
\begin{minipage}[b]{.48\linewidth}
  \centering
  \centerline{\includegraphics[width=\linewidth]{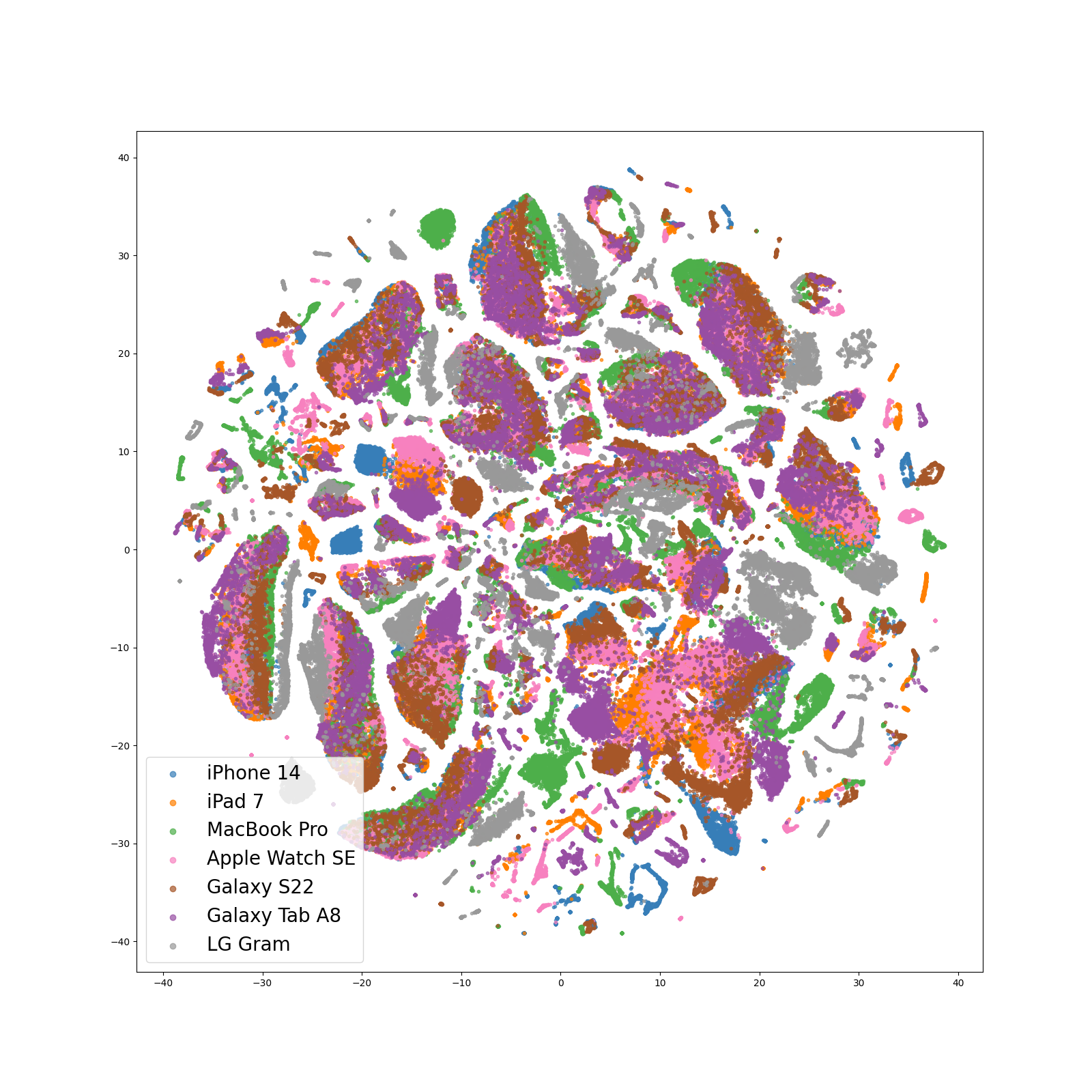}}
  \subcaption{All sound (To iPhone 14)}
  \label{fig:tsne(c)}
\end{minipage}
\hfill
\begin{minipage}[b]{0.48\linewidth}
  \centering
  \centerline{\includegraphics[width=\linewidth]{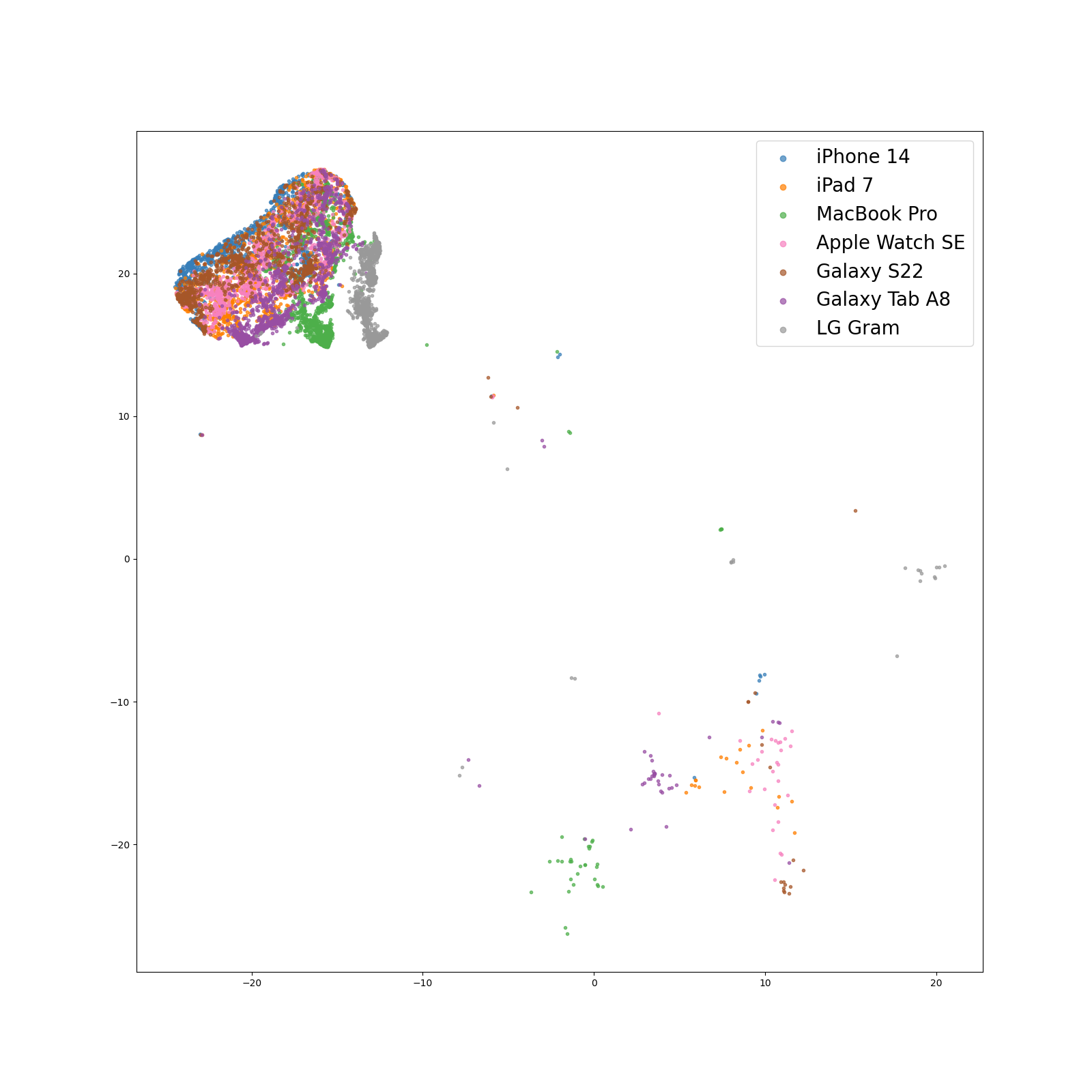}}
  \subcaption{Whistle (To iPhone 14)}
  \label{fig:tsne(d)}
\end{minipage}
\caption{2D t-SNE\cite{JMLR:v9:vandermaaten08a} visualization of intermediate embeddings of the baseline ResNet50 on the development data. (a) and (b) illustrate the full development set and whistle sound samples drawn from it, respectively. (c) and (d) are sound samples converted to iPhone 14 using (a) and (b).}
\label{fig:tsne_2d}
\end{figure}

\subsection{Network Architecture}
\label{section:network_architecture}
We adapt the CycleGAN implementation from the original author's code\footnote{https://github.com/junyanz/pytorch-CycleGAN-and-pix2pix}. Our generator includes two up- and down-sampling layers and nine residual blocks with instance normalization, as in \cite{zhu2017unpaired}. The tanh layer is omitted. For the discriminator, we employ a 16x16 \emph{Patch}GAN\cite{isola2017image} with instance normalization for better convergence and to minimize blurriness in the output.

\begin{table*}[t]
\caption{Results for generalization/adaptation capability of previous methods and our methods on the validation set. Source device(S) is iPhone 14, and target devices(T1$\sim$T6) are Galaxy S22, iPad 7, Galaxy Tab A8, Apple Watch SE, Macbook Pro('20), and LG Gram('20), respectively. The last column shows an average and 95\% confidence interval of the performance.}
\label{tab:result}
\centering
\begin{tabular}{lcccccccc}
\toprule
\multicolumn{1}{c}{\multirow{2}{*}{Method}} & \multicolumn{8}{c}{F1 Score}                                                                                     \\ \cmidrule{2-9} 
\multicolumn{1}{c}{}                        & S & T1 & T2 & T3 & T4 & T5 & T6 & Overall (- S) \\ \midrule
Baseline                                    & 0.982     & 0.409      & 0.709  & 0.248         & 0.471          & 0.687       & 0.491   & 0.503 ± 0.167         \\
Gaussian Noise                              & 0.983     & 0.708      & 0.918  & 0.576         & 0.565          & 0.780       & 0.683   & 0.705 ± 0.127         \\
Reverberation                               & 0.980     & 0.895      & 0.852  & 0.539         & 0.832          & 0.736       & 0.360   & 0.702 ± 0.202         \\
Pitch Shift                                 & 0.981     & 0.471      & 0.744  & 0.221         & 0.658          & 0.648       & 0.442   & 0.531 ± 0.183         \\
SpecAugment                                 & \textbf{0.985}     & 0.372      & 0.762  & 0.214         & 0.363          & 0.634       & 0.324   & 0.445 ± 0.199         \\
MixUp                                       & 0.983     & 0.336      & 0.677  & 0.213         & 0.449          & 0.656       & 0.387   & 0.453 ± 0.175         \\
FilterAugment                               & 0.981     & 0.964      & 0.891  & 0.586         & 0.874          & 0.794       & 0.642   & 0.792 ± 0.143         \\
Freq-MixStyle                               & 0.974     & 0.839      & 0.879  & 0.795         & 0.902          & \textbf{0.885}       & 0.832   & 0.855 ± 0.038         \\
RFN                                         & 0.980     & 0.919      & 0.909  & 0.742         & 0.907          & 0.829       & 0.614   & 0.820 ± 0.116         \\
MC-100-Gen                          & 0.981     & 0.958      & \textbf{0.912}  & 0.894         & 0.899          & 0.831       & 0.852   & 0.891 ± 0.043         \\
MC-200-Gen                          & 0.982     & \textbf{0.969}      & 0.909  & \textbf{0.903}         & \textbf{0.912}          & 0.859       & \textbf{0.887}   & \textbf{0.907 ± 0.035}         \\ \midrule
MC-100-Adapt (p=0.5) & - & 0.956 & 0.905 & 0.904 & 0.880 & 0.902 & 0.890 & 0.906 ± 0.025 \\
MC-100-Adapt (p=1.0) & - & 0.891 & 0.906 & 0.855 & 0.834 & 0.811 & 0.808 & 0.851 ± 0.039 \\
MC-200-Adapt (p=0.5) & - & 0.965 & \textbf{0.922} & \textbf{0.906} & 0.910 & \textbf{0.908} & \textbf{0.907} & \textbf{0.920 ± 0.022} \\
MC-200-Adapt (p=1.0) & - & 0.935 & 0.917 & 0.864 & 0.873 & 0.880 & 0.848 & 0.886 ± 0.032 \\ \midrule
Real                                        & 0.983     & 0.982      & 0.972  & 0.985         & 0.979          & 0.983       & 0.986   & 0.981 ± 0.005         \\ \bottomrule
\end{tabular}
\end{table*}

\subsection{Visualization of Generated Samples}
Figure \ref{fig:real_gen_spectrogram} demonstrates that the Microphone Conversion network accurately replicates the spectro-temporal characteristics of each target device. We visualize difference spectra between devices using Welch's method, and Figure \ref{fig:impulse_response} confirms the network's ability to capture each device's unique frequency response. We average spectrogram samples along with temporal dimension to derive spectra of different devices. Feature-level analysis, shown in Figure \ref{fig:tsne_2d}, validates that our method produces outputs closely aligned with the target device, establishing its efficacy for data augmentation.

\begin{table}[]
\caption{Results for the second adaptation scenario. Training Set 1 and 2 are (GalaxyS22, LG Gram, GalaxyTabA8) and (iPad7, AppleWatchSE, MacbookPro), respectively.}
\label{tab:adaptation_result2}
\centering
\begin{tabular}{lcl}
\toprule
\multicolumn{1}{c}{\multirow{2}{*}{Method}} & \multicolumn{2}{c}{F1 Score}          \\ \cmidrule{2-3} 
\multicolumn{1}{c}{}                        & Set 1         & Set 2                 \\ \midrule
Baseline                                    & 0.917         & 0.975                 \\
MC-100-Adapt (p=0.5)                        & 0.976         & \textbf{0.978}        \\
MC-100-Adapt (p=1.0)                        & 0.940         & 0.971                 \\
MC-200-Adapt (p=0.5)                        & 0.977         & \textbf{0.978}        \\
MC-200-Adapt (p=1.0)                        & \textbf{0.979}& \textbf{0.978}        \\ \midrule
Real                                        & 0.983         & 0.983                 \\ \bottomrule
\end{tabular}
\end{table}

\section{Experiments and Results}
In this section, we explore the performance decline in SEC systems when faced with heterogeneous recording devices. We demonstrate that incorporating Microphone Conversion can significantly improve the SEC systems' resilience to device variability. Additionally, we show our approach is effective in adapting SEC systems for specific devices.

\subsection{Implementation Details}
\label{section:implementation}
The audio data, down-sampled to 22,050 Hz, is converted into log Mel spectrograms using a 1,024-sample Hanning window, a 256-sample hop length, and 80 Mel bands. We replace the negative log likelihood loss in Eq. (\ref{eq:gan_loss}) with the least square loss\cite{mao2017least} for Microphone Conversion models. we randomly draw an initial learning rate from [0.00002, 0.002] and halved at random intervals between [10, 50], optimized by the Adam optimizer with $\beta=(0.5, 0.999)$, with a batch size of 100. Discriminator updates employ a generated image buffer\cite{shrivastava2017learning}. These models are trained for 100 and 200 epochs, as MC-100 and MC-200, using $data_{train,mc}$, respectively. We set $\lambda$ to 10 for ${L}_{cycle}$ in the total loss function to control its relative importance. Parameter tuning is done using 10 iterations of Bayesian search, minimizing $\mathcal{L}_{cycle}$ for final evaluation.

For SEC systems, we employ ResNet50\cite{he2016residual} as our baseline network and use iPhone 14 segments of \emph{$data_{train,sec}$} for training. We specifically choose the iPhone 14 as our source device because smartphones are among the most commonly used devices for sound reception in the modern world, making them ideal for our study. Model optimization is carried out using AdamW\cite{loshchilov2018decoupled} with $\beta=(0.9, 0.999)$ and an initial learning rate of 0.001, reduced by 0.1 every 25 epochs. All implementations are executed on NVIDIA T4 and RTX6000 GPUs. During training, the source data, iPhone 14, is randomly converted to six other devices through Microphone Conversion networks to assess generalization capabilities. MC-100-Gen and MC-200-Gen correspond to this strategy in Table \ref{tab:result}.

We also evaluate our methods in two adaptation scenarios, one with a single source (iPhone 14) and target, and another with multiple sources and a single target (iPhone 14), with probabilities set at 0.5 and 1.0 for applying Microphone Conversion. MC-100-Adapt and MC-200-Adapt in Table \ref{tab:result} and \ref{tab:adaptation_result2}, respectively, correspond to the single-target and multi-target strategy. Finally, a `Real' performance metric is introduced as an ideal case, where training and inference device matches.

\subsection{Recent Approaches}
\label{section:recent_approaches}
We examine top-performing methods from recent DCASE Challenges that mitigate device variability. Freq-MixStyle\cite{Schmid2022} and Residual Normalization\cite{Kim2021b} utilize frequency-wise statistics for better generalization. An extended version, Relaxed Instance Frequency-wise Normalization (RFN)\cite{kim22_interspeech}, is applied after each bottleneck block, resulting in 5 RFN blocks. FilterAugment\cite{Nam2021} adjusts spectrograms via randomly generated filters, and we apply its optimal linear type\cite{9747680}. We also test standard data augmentations like Gaussian noise, room impulse response, pitch shift, SpecAugment\cite{Park2019}, and MixUp\cite{zhang2018mixup}, all with a 0.5 probability, except for RFN.

\subsection{Results}
Our evaluation, summarized in Tables \ref{tab:result} and \ref{tab:adaptation_result2}, reveals SEC system performance vulnerabilities when faced with heterogeneous recording devices. For instance, there is up to a 73.7\% performance drop compared to the `Real' metric in the case of T3. However, models perform better on devices like T2 and T5 that share similar features with the training data, as indicated by their closeness in the t-SNE feature space, showing a modest performance drop of 26.3\% and 29.5\%.

Traditional data augmentations provide limited improvement. Among recent approaches, Freq-MixStyle is most effective, achieving an average F1 score of 85.5\% with a 3.8\% confidence interval. MC-200-Gen (Ours), excels with an average F1 score of 90.7\% and minimal variability (3.5\%).

In adaptation scenarios, SEC models using an adaptation strategy outperform MC-200-Adapt (p=0.5) on 4 out of 6 targets, gaining 1.3\% in overall performance. They also nearly match the `Real' performance metric when trained on diverse devices. Longer training for our Microphone Conversion networks boosts both generalization and adaptability, hinting at the potential for further optimization.

\section{Conclusions and Limitations}
To tackle device variability in SEC systems, we developed a specialized sound event dataset recorded across multiple real-world devices in an anechoic chamber. Our introduced Microphone Conversion method significantly improves SEC performance, outperforming recent approaches in both generalization and adaptation tasks. However, the CycleGAN component of our solution assumes a one-to-one domain mapping, requiring separate models for each domain pair. Future work could explore integrating impulse response with CycleGAN for more versatile domain mapping.

\vfill\pagebreak

\bibliographystyle{IEEEbib}
\bibliography{mic-conversion}

\end{document}